\documentclass[11pt]{article} 
\usepackage{graphicx,type1cm,eso-pic,color}
\usepackage{amsfonts, amssymb}
\makeindex

\newtheorem{theorem}{Theorem}
 
\newtheorem{definition}[theorem]{Definition} 
\newtheorem{corollary}[theorem]{Corollary} 
\newtheorem{remark}[theorem]{Remark}

\title{\bf Huygens' Principle  for the Klein-Gordon equation in the de~Sitter spacetime}

\author{Karen Yagdjian \\
{}\\
\small Department of Mathematics,
\small University of Texas-Pan American,\\
\small 1201 W.~University Drive, Edinburg, TX 78539,
\small USA, \\ 
\small {yagdjian@utpa.edu}}

\begin{document}
\date{}
\maketitle

\thispagestyle{empty} 
\vspace{-0.3cm}

\begin{abstract}  \small  
In this article we   prove that the   Klein-Gordon equation in the de Sitter spacetime obeys the Huygens' principle only if the physical mass $m$ of the scalar field and  the dimension $n\geq 2$ of the spatial variable are tied by the equation $m^2=(n^2-1)/4 $. Moreover,   we define the incomplete Huygens' principle, which is the Huygens' principle
restricted to the   vanishing second initial datum,   and then reveal that  the massless     scalar field  in the de Sitter spacetime
obeys the   incomplete Huygens' principle and does not obey  the  Huygens' principle, for the dimensions $n=1,3$, only. 
Thus, in the de~Sitter spacetime the existence of two different scalar fields (in fact, with $m=0$ and $m^2=(n^2-1)/4 $), 
which obey  incomplete Huygens' principle, 
is equivalent  to the condition $n=3$ (in fact, the spatial dimension of the physical world).
For $n=3$ these two values of the mass are the endpoints of the so-called in  quantum field theory the Higuchi bound.
The value  $m^2=(n^2-1)/4 $  of the physical mass allows us also to obtain complete asymptotic expansion of the solution for the large time.\\
{}  
\end{abstract}

\section{Introduction and Statement of Results}
\label{sec:1}

In this article we    prove that the   Klein-Gordon equation in the de Sitter spacetime obeys the Huygens' principle only if the physical mass $m$ of the scalar field and  the dimension $n\geq 2$ of the spatial variable are tied by the equation $m^2=(n^2-1)/4 $. Moreover,   we define the incomplete Huygens' principle, which is the Huygens' principle
restricted to the   vanishing second initial datum,   and then reveal that  the massless     scalar field  in the de Sitter spacetime
obeys the   incomplete Huygens' principle and does not obey  the  Huygens' principle, for the dimensions $n=1,3$, only.  
\smallskip

The Klein-Gordon equation arising in relativistic physics and, in particular, general relativity and cosmology, as well as, in more recent  quantum field theories,
is a covariant equation that is considered in the curved pseudo-Riemannian
manifolds.  (See, e.g., Birrell and Davies~\cite{Birrell-Davies}, Parker and Toms~\cite{{Parker-Toms}}, Weinberg~\cite{Weinberg}.) Moreover,
the latest astronomical  observational discovery that the expansion of the universe is speeding   supports the model of the expanding universe that
is mathematically described by the manifold with  metric tensor depending on time and spatial variables. In this paper we restrict ourselves to the manifold arising in the
so-called de~Sitter   model of the universe, which is the curved manifold due to the cosmological constant.
\smallskip

The  line element in the de~Sitter spacetime has  the form
\begin{eqnarray}
\label{ds}
ds^2= - \left( 1- \frac{r^2}{R^2}\right) c^2\, dt^2+ \left( 1- \frac{r^2}{R^2}\right)^{-1}dr^2 + r^2(d\theta ^2 + \sin^2 \theta \, d\phi ^2)\,.
\end{eqnarray}
The  Lama{\^i}tre-Robertson transformation
$
r'=\frac{r}{\sqrt{1-r^2/R^2}} e^{-ct/R}$, \, $ t'=t+\frac{R}{2c} \ln \left( 1- \frac{r^2}{R^2}\right) $, \, $ \theta '=\theta $, \, $ \phi '=\phi \,, 
$
leads to the following form for the line element \cite[Sec.134]{Moller}, \cite[Sec.142]{Tolman}:
$
ds^2= -   c^2\, d{t'}^2+e^{2ct'/R}( d{r'}^2   + r'^2\,d{\theta'} ^2 + r'^2\sin^2 \theta' \, d{\phi '}^2)\,.
$
Finally, defining new space coordinates $x'$, $y'$, $z'$ connected with $r'$, $\theta '$, $\phi '$
by the usual equations connecting Cartesian  coordinates and polar coordinates in a Euclidean space, (\ref{ds}) may be written 
\cite[Sec.134]{Moller}
\[
ds^2= -   c^2\, d{t'}^2+ e^{2ct'/R}( d{x'}^2   + d{y'} ^2 +  d{z '}^2)\,.
\]
The new coordinates $x'$, $y '$, $z '$, $t'$ can take all values from $-\infty$ to $\infty$.  Here $R$ is the ``radius'' of the universe. In fact, the de~Sitter model belongs to the family of  the Friedmann-Lema{\^i}tre-Robertson-Walker spacetimes (FLRW spacetimes).
In the FLRW  spacetime \cite{Hawking}, one can choose coordinates so that the metric has the form $
ds^2=-dt^2+a^2(t)d \sigma ^2$.

\smallskip

The homogeneous and isotropic cosmological models possess the highest degree of symmetry that makes them  more amenable to rigorous study.
Among them we mention FLRW  models. The simplest class of cosmological models
can be obtained if we assume, additionally, that the metric of the slices of constant time is flat and that
the spacetime metric can be written in the form
$
ds^2= -dt^2+ a^2(t)( d{x}^2   + d{y} ^2 +  d{z}^2 )
$
with an appropriate scale factor $a(t)$.
The assumption that the universe is expanding leads to the positivity of
  the time derivative $\frac{d }{dt}a (t)$. A further assumption that the universe obeys the accelerated expansion suggests
that the second derivative  $\frac{d^2 }{dt^2}a (t)$  is positive.
Under the assumption of FLRW symmetry the equation of motion in the case of positive
cosmological constant \, $\Lambda $ \, leads to the solution
$a(t)=a(0)e^{t\sqrt{\frac{\Lambda }{3}}}$, 
which produces models with exponentially accelerated expansion, which is  referred to as  the {\it de Sitter
model}.
\smallskip

In  quantum field theory    the matter fields are described by the function  $\phi $  must satisfy  equations of motion.  
In the case of the massive scalar field, the equation of motion is   the  Klein-Gordon equation generated by the metric $g$:
\[
\frac{1}{\sqrt{|g|}}\frac{\partial }{\partial x^i}\left( \sqrt{|g|} g^{ik} \frac{\partial \phi  }{\partial x^k} \right) = m^2 \phi  + V'(\phi ) \,.
\]
In physical
terms this equation describes a local self-interaction for a scalar particle.
In the  de~Sitter universe  the equation for the scalar field with mass \,  $m$\,   and potential function \, $V$   \,
written out explicitly in coordinates is 
\begin{equation}
\label{1.3}
  \phi_{tt} +   n H \phi_t - e^{-2Ht} \bigtriangleup \phi + m^2\phi=   - V'(\phi )\,.
\end{equation}
Here $x \in {\mathbb R}^n$, $t \in {\mathbb R}$, and $\bigtriangleup $ is
 the Laplace operator on the flat metric, $\bigtriangleup := \sum_{j=1}^n \frac{\partial^2 }{\partial x_j ^2} $,
while $H = \sqrt{\Lambda /3}$ is the Hubble constant. For the sake of simplicity, henceforth, we set $H=1$. A typical example of a potential function  would be  $V(\phi )=\phi ^4$.
\smallskip

For the solution $\Phi$ of the Cauchy problem for the linear Klein-Gordon equation 
\begin{equation}
\label{1.10}
  \Phi_{tt} +   n   \Phi_t - e^{-2 t} \bigtriangleup \Phi + m^2\Phi=  0\,, \quad \Phi (x,0)= \varphi_0 (x)\, , \quad \Phi _t(x,0)=\varphi_1 (x)\,,
\end{equation}
the following formula is obtained in \cite{Yag_Galst_CMP}:
\begin{eqnarray}
\label{large}
&  &
\Phi (x,t) \\
& = &
e^{-\frac{n-1}{2}t} v_{\varphi_0}  (x, \phi (t))
+ \,  e^{-\frac{n}{2}t}\int_{ 0}^{1} v_{\varphi_0}  (x, \phi (t)s)\big(2  K_0(\phi (t)s,t)+ nK_1(\phi (t)s,t)\big)\phi (t)\,  ds  \nonumber \\
& &
+\, 2e^{-\frac{n}{2}t}\int_{0}^1   v_{\varphi _1 } (x, \phi (t) s)
  K_1(\phi (t)s,t) \phi (t)\, ds
, \quad x \in {\mathbb R}^n, \,\, t>0\,, \nonumber 
\end{eqnarray}
provided that the mass $m$ is large, that is,  $m^2 \geq n^2/4 $.  Here, $\phi (t):= 1-e^{-t} $ and   for $x \in {\mathbb R}^n$,
the function $v_\varphi  (x, \phi (t) s)$  coincides with the value $v(x, \phi (t) s) $
of the solution $v(x,t)$ of the Cauchy problem
\begin{equation}
\label{1.11} 
v_{tt}-  \bigtriangleup v =0, \quad v(x,0)= \varphi (x), \quad v_t(x,0)=0\,.
\end{equation}
To define the kernels $K_0(z,t) $ and $K_1(z,t) $ we  introduce the following notations.
First, we define a  {\it chronological future}  
$D_+ (x_0,t_0) $ of the point  (event) $(x_0 ,t_0)$,   $x_0 \in {\mathbb R}^n$, $t_0 \in {\mathbb R}$,
and  a  
{\it chronological past} $D_- (x_0,t_0) $  of the point (event) $(x_0 ,t_0)$,  $x_0 \in {\mathbb R}^n$, $t_0 \in {\mathbb R}$,
 as follows
\begin{eqnarray*} 
D_\pm (x_0,t_0)
& := &
\Big\{ (x,t)  \in {\mathbb R}^{n+1}  \, ; \,
|x -x_0 | \leq \pm( e^{-t_0} - e^{-t })
\,\Big\} \,.
\end{eqnarray*}
In fact,  any intersection of  $ D_- (x_0,t_0) $ with the hyperplane $t=const <t_0$ determines the so-called dependence domain
for the point $(x_0,t_0) $, while the  intersection of  $ D_+ (x_0,t_0) $
with the hyperplane $t=const >t_0$ is the so-called  domain of influence of the point $(x_0,t_0) $.  
We define also the {\it characteristic conoid} (ray cone) by
\begin{eqnarray*} 
C_\pm (x_0,t_0)
& := &
\Big\{ (x,t)  \in {\mathbb R}^{n+1}  \, ; \,
|x -x_0 | = \pm( e^{-t_0} - e^{-t })
\,\Big\} \,.
\end{eqnarray*}
Thus,  the characteristic conoid  $C_+ (x_0,t_0) $ ($C_- (x_0,t_0) $) is the surface of the chronological future $D_+ (x_0,t_0)$ (chronological past $D_+ (x_0,t_0)$)
of the  point   $(x_0 ,t_0)$.

Then, we define for $(x_0, t_0) \in {\mathbb R}^n\times R$  the function
\begin{eqnarray}
\label{E}
E(x,t;x_0,t_0)
& =  &
(4e^{-t_0-t })^{iM} \Big((e^{-t }+e^{-t_0})^2 - (x - x_0)^2\Big)^{-\frac{1}{2}-iM    } \\
&  &
F\Big(\frac{1}{2}+iM   ,\frac{1}{2}+iM  ;1;
\frac{ ( e^{-t_0}-e^{-t })^2 -(x- x_0 )^2 }{( e^{-t_0}+e^{-t })^2 -(x- x_0 )^2 } \Big)\nonumber
\end{eqnarray}
in    $D_+ (x_0,t_0)\cup D_- (x_0,t_0) $,  where $F\big(a, b;c; \zeta \big) $ is the hypergeometric function. (For the definition of $F\big(a, b;c; \zeta \big) $ see, e.g., \cite{B-E}.) Here the notation $x^2=  x\cdot x =|x|^2$ for  $ x   \in {\mathbb R}^n  $  has been used.
The kernels  $K_0(z,t)   $    and $K_1(z,t)   $ are defined by
\begin{eqnarray}
\label{K0}
&  &
K_0(z,t)
  :=
- \left[  \frac{\partial }{\partial b}   E(z,t;0,b) \right]_{b=0} \\
&  = &
 (4e^{-t })^{iM} \big((1+e^{-t })^2 - z^2\big)^{ -iM    } \frac{1}{ [(1-e^{ -t} )^2 -  z^2]\sqrt{(1+e^{-t } )^2 - z^2} }
\nonumber \\
&   &
\times  \Bigg[  \big(  e^{-t} -1 - iM(e^{ -2t} -      1 -  z^2) \big)
F \Big(\frac{1}{2}+iM   ,\frac{1}{2}+iM  ;1; \frac{ ( 1-e^{-t })^2 -z^2 }{( 1+e^{-t })^2 -z^2 }\Big)\nonumber  \\
&  &
\hspace{0.3cm}  +   \big( 1-e^{-2 t}+  z^2 \big)\Big( \frac{1}{2}-iM\Big)
F \Big(-\frac{1}{2}+iM   ,\frac{1}{2}+iM  ;1; \frac{ ( 1-e^{-t })^2 -z^2 }{( 1+e^{-t })^2 -z^2 }\Big) \Bigg]\nonumber 
\end{eqnarray}
and $K_1(z,t)  :=    E(z ,t;0,0) $, that is,
\begin{eqnarray}
\label{K1}
&  &
K_1(z,t)  \\
 & = &
 (4e^{ -t })^{iM} \big((1+e^{-t })^2 -   z  ^2\big)^{-\frac{1}{2}-iM    }
F\left(\frac{1}{2}+iM   ,\frac{1}{2}+iM  ;1;
\frac{ ( 1-e^{-t })^2 -z^2 }{( 1+e^{-t })^2 -z^2 } \right), \nonumber \\
&  &
\hspace{4cm}  0\leq z\leq  1-e^{-t},\nonumber 
\end{eqnarray}
respectively. Here $M=\sqrt{m^2-\frac{n^2}{4}} $. The main properties of $K_0(z,t)  $ and $K_1(z,t)  $  are listed and proved in Section~3~\cite{Yag_Galst_CMP}.
\smallskip

For the case of small mass, $m^2 \leq n^2/4$, the similar formula is obtained in \cite{yagdjian_DCDS}. More precisely,
if we denote  $M=\sqrt{\frac{n^2}{4}-m^2} $, then for the solution $\Phi $ of the Cauchy problem  (\ref{1.10}), there is a representation
\begin{eqnarray}
\label{24}
\Phi  (x,t)
& = &
e^{-\frac{n-1}{2}t} v_{\varphi_0}  (x, \phi (t))\\
&  &
+ \,  e^{-\frac{n}{2}t}\int_{ 0}^{1} v_{\varphi_0}  (x, \phi (t)s)\big(2  K_0(\phi (t)s,t;M)+ nK_1(\phi (t)s,t;M)\big)\phi (t)\,  ds  \nonumber \\
& &
+\, 2e^{-\frac{n}{2}t}\int_{0}^1   v_{\varphi _1 } (x, \phi (t) s)
  K_1(\phi (t)s,t;M) \phi (t)\, ds
, \quad x \in {\mathbb R}^n, \,\, t>0\,. \nonumber 
\end{eqnarray}
 Here we have used the new functions $E(x,t;x_0,t_0;M) $, $K_0(z,t;M)   $,    and $K_1(z,t;M) $, which can be obtained by the analytic continuation
of  the functions $E(x,t;x_0,t_0) $, $K_0(z,t)   $,    and $K_1(z,t) $, respectively, to the  complex domain.
 First we define the function
\begin{eqnarray}
\label{EM}
E(x,t;x_0,t_0;M)
& =  &
 4 ^{-M}  e^{ M(t_0+t) } \Big((e^{-t }+e^{-t_0})^2 - (x - x_0)^2\Big)^{-\frac{1}{2}+M    } \\
 &  &
\times F\Big(\frac{1}{2}-M   ,\frac{1}{2}-M  ;1;
\frac{ ( e^{-t_0}-e^{-t })^2 -(x- x_0 )^2 }{( e^{-t_0}+e^{-t })^2 -(x- x_0 )^2 } \Big) . \nonumber
\end{eqnarray}
Hence, it is related to the function $E(x,t;x_0,t_0) $ of (\ref{E}) as follows:
\[
E(x,t;x_0,t_0)=E(x,t;x_0,t_0;-iM)\,.
\]
Next
we define also new kernels  $K_0(z,t;M)   $    and $K_1(z,t;M) $ by
\begin{eqnarray}
\label{K0M}
&  &
K_0(z,t;M)
  :=
- \left[  \frac{\partial }{\partial b}   E(z,t;0,b;M) \right]_{b=0} \\
&  = &
4 ^ {-M}  e^{ t M}\big((1+e^{-t })^2 - z^2\big)^{  M    } \frac{1}{ [(1-e^{ -t} )^2 -  z^2]\sqrt{(1+e^{-t } )^2 - z^2} }
\nonumber \\
&   &
\times  \Bigg[  \big(  e^{-t} -1 +M(e^{ -2t} -      1 -  z^2) \big)
F \Big(\frac{1}{2}-M   ,\frac{1}{2}-M  ;1; \frac{ ( 1-e^{-t })^2 -z^2 }{( 1+e^{-t })^2 -z^2 }\Big)\nonumber  \\
&  &
\hspace{0.3cm}  +   \big( 1-e^{-2 t}+  z^2 \big)\Big( \frac{1}{2}+M\Big)
F \Big(-\frac{1}{2}-M   ,\frac{1}{2}-M  ;1; \frac{ ( 1-e^{-t })^2 -z^2 }{( 1+e^{-t })^2 -z^2 }\Big) \Bigg]\nonumber 
\end{eqnarray}
and $K_1(z,t;M)   :=
  E(z ,t;0,0;M) $, that is,
\begin{eqnarray}
\label{K1M}
&  &
K_1(z,t;M)  \\
& = &
  4 ^{-M} e^{ Mt }  \big((1+e^{-t })^2 -   z  ^2\big)^{-\frac{1}{2}+M    }
F\left(\frac{1}{2}-M   ,\frac{1}{2}-M  ;1;
\frac{ ( 1-e^{-t })^2 -z^2 }{( 1+e^{-t })^2 -z^2 } \right), \nonumber \\
&  &
\hspace{5cm}  0\leq z\leq  1-e^{-t},\nonumber 
 \end{eqnarray}
respectively.  In fact, $E(x,t;x_0,t_0;M) $ coincides with $E(x,t;x_0,t_0) $ if we replace $ M$ with $iM $, that is, 
it is an analytic continuation of the function $E(x,t;x_0,t_0) $ to the complex 
plane $M \in {\mathbb C}$.   The same statement is true for the functions $K_0(z,t;M)   $
and $K_1(z,t;M)   $.
\smallskip

The expressions (\ref{large}) and (\ref{24}) can be regarded as the integral transforms applied to the solution of  (\ref{1.11}). (See for details \cite{Rendi_Trieste}.) 
According to \cite{Rendi_Trieste}, the fundamental solutions (the retarded and advanced  Green functions) of the operator have the similar representations.
\smallskip

Suppose now that we are looking for the simplest possible kernels $K_0(z,t;M)   $    and \\
$K_1(z,t;M) $   of the   integral transforms. 
Surprisingly that perspective shades a light on the quantum field theory in the de~Sitter universe   and 
 reveals a new unexpected  link between   the 
Higuchi bound \cite{Higuchi} and the Huygens' principle. 
\smallskip

 Indeed, in the hierarchy of the  hypergeometric functions
the simplest one is the constant, $F \left(0, 0;1; \zeta  \right) =1$.  The parameter $M$ leading to  such function  $F \left(0, 0;1; \zeta  \right) = 1  $ is $M=\frac{1}{2}$, and, consequently, $m^2= \frac{n^2-1}{4} $. 
\smallskip

The next simple function of that  hierarchy is a linear function. That  function $F \left(a, b;1; \zeta  \right) $ has the parameters $a=b=-1$
and coincides with the polynomial $1+ \zeta $. The parameter $M$ leading to  such function  $F \left(-1, -1;1; \zeta  \right) = 1+ \zeta $ is     $M=\frac{3}{2} $, 
and, consequently, $m^2= \frac{n^2-9}{4} $.   
\smallskip

In the case of $n=3$ the only real masses, which simplify the kernels, that is, make $F $ polynomial,    are given by  $M=\frac{1}{2}$ and $M=\frac{3}{2}$. For the  square of the physical mass $ m^2$ they are  $m^2=2$ and  $m= 0$, respectively.  These are exactly the endpoints of the 
interval $(0,2) $ that,  in the case of $n=3 $, is known as  the so-called Higuchi bound \cite{Higuchi}. In the physical variables it is the interval $(0, 2\Lambda /3) $.
\smallskip

It turns out that the interval $  ( 0, \sqrt{2}  ) $     plays significant role   in the linear quantum field theory ~\cite{Higuchi}, in completely different context than the explicit representation of the solutions of the Cauchy problem.
More precisely,   the   Higuchi bound~\cite{Higuchi},\cite{Deser-Waldron},\cite{Lasma Alberte},\cite{Berkhahn-Dietrichb-Hofmann},\cite{Dengiz-Tekin} 
   arises  in the quantization of free massive fields with the spin-2 in the de~Sitter spacetime with $n=3$.
It is the forbidden mass range for spin-$2$ field theory in de~Sitter spacetime because of the appearance of negative norm states. 
Thus, the point $m= \sqrt{2} $ is exceptional for the the quantum fields theory in the de~Sitter spacetime.  In particular, for massive spin-2 fields, it is known \cite{Deser-Waldron}, \cite{Higuchi} that the norm of the helicity zero mode changes sign across the line $m^2= 2 $. The region  $m^2<2$ is therefore unitarily forbidden. It is noted in \cite{Lasma Alberte} that all canonically normalized helicity $-0,\pm 1,\pm 2$ modes of massive graviton on the de Sitter universe satisfy Klein-Gordon equation for a {\it massive scalar field with the same effective mass}.  Then, it is known (see, e.g., \cite{Deser-Nepomechie}) that, if $m^2=2$, the action 
is invariant under the gauge  transformation, and that invariance already suggests that there exists some discontinuity in the theory at $m^2=2$.
\smallskip

In the case of   $n\in {\mathbb N}$ we obtain for the physical mass several points,  $m^2= \frac{n^2}{4}-\left( \frac{1}{2} +k\right)^2$, $k=0,1,\ldots,\left[\frac{n-1}{2}\right] $,
which make $F$ polynomial.
We will call these points {\it knot points} for the mass of the equation. For $n=1$  only  the massless field $m=0$ has knot point. 
\smallskip

The explicit representation formulas allows us to prove in Section~\ref{SS3.1cr} that the  largest knot point, and, in particular, the right endpoint of the Higuchi bound if $n=3$, is the only value of the mass of the particle which produces 
scalar field that obeys the Huygens' principle.  Recall (see, e.g., \cite{Gunther}) that a hyperbolic equation is said to satisfy Huygens'
principle if the solution vanishes at all points which cannot be
reached from the support of initial data by a null geodesic, that is, there is
no tail. The tails are important within cosmological context.(See, e.g., \cite{Ellis-Sciama},\cite{Gleiser-Price},\cite{FARAONI-GUNZIG} and references therein.) 
\smallskip

An exemplar equation satisfying Huygens' principle is the
wave equation in $n + 1$ dimensional Minkowski spacetime for odd $n
\geq  3$. According to Hadamard's conjecture (see, e.g., \cite{Gunther,Berest,{Ibragimov-Oganesyan}}) this is the only
(modulo transformations of coordinates and unknown function)
huygensian linear second-order hyperbolic equation. There exists an extensive literature on the Huygens' principle in the 4-dimensional  spacetime of constant curvature (see e.g. \cite{FARAONI-GUNZIG},\cite{Sonego-Faraoni} and references therein). 
\smallskip

In the present article we have  a new proof of the following theorem.
\begin{theorem}
\label{T3}
The value $m= \sqrt{n^2-1}/2$ is the only value of the physical mass $m $, such that the solutions of the equation 
\begin{eqnarray}
\label{25}
  \Phi _{tt} +   n   \Phi _t - e^{-2 t} \bigtriangleup \Phi  + m^2\Phi = 0,
\end{eqnarray}
 obey the  Huygens' principle,
whenever  the wave equation in the  Minkowski spacetime does, that is, $n \geq 3 $ is an odd number.
\end{theorem}

\medskip

Even if the equation is not huygensian (not tail-free for some admissible data), one might nevertheless be interested in data that produce tail-free solution.
Such data are prescribed in the following definition which is hinted by the string equation.  
\begin{definition}
We say that the equation obeys the  incomplete   Huygens' principle with respect to the first initial datum,
   if the solution with the second datum  $\varphi _1 =0 $ vanishes at all points which cannot be
reached from the support of initial data by a null geodesic.
\end{definition}

If the equation  obeys the   Huygens' principle, then it obeys also the  incomplete   Huygens' principle with respect to the first initial datum.
However, the equation in the de~Sitter spacetime   shows that the converse is not true. 
\begin{theorem}
\label{TIHP}
Suppose that equation (\ref{25}) does not obey the   Huygens' principle. Then,  it obeys the  incomplete   Huygens' principle with respect to the first initial datum,  if and only if the equation 
is massless, $m=0$, and  either $n=1 $  or $n= 3$.
\end{theorem}

We have to point out that for the classical string equation, the  Huygens' principle is not valid, but the D'Alembert formula shows that incomplete   Huygens' principle with respect to the first initial datum is  fulfilled. 
By combining Theorem~\ref{T3} and Theorem~\ref{TIHP} we arrive at the following interesting conclusion.

\begin{corollary}

Assume that the equations $
  \Phi _{tt} +   n   \Phi _t - c_1^2 e^{-2 t} \bigtriangleup \Phi  + m_1^2\Phi = 0$ 
and 
 $ \Phi _{tt} +   n   \Phi _t - c_2^2 e^{-2 t} \bigtriangleup \Phi  + m_2^2\Phi = 0 $, where $c_1$, $c_2$ are positive numbers,   
obey     the  incomplete  Huygens' principle. Then they describe the fields with different mass, $m_1\not=m_2$, (in fact, $ \frac{\sqrt{n^2-1}}{2}$
and $ 0$) if and only if the dimension $n$  of the spatial variable $x$ is $3$.
\end{corollary}

Thus, in the de~Sitter spacetime the existence of two different scalar fields (in fact, with $m=0$ and $m^2=(n^2-1)/4 $), 
which obey  incomplete Huygens' principle, 
is equivalent  to the condition $n=3$.
The  dimension  $n=3$ of the last corollary agrees with the experimental data.

\medskip

This paper is organized as follows.  In Section \ref{S2} we define the incomplete Huygens' principle. Then, 
 in Theorem~\ref{TIHP}, we give  description of the class of equations which obey 
that principle. The proofs of Theorem~\ref{T3} and Theorem~\ref{TIHP} are given in Section~\ref{SS3.1cr}. 
For the value $m= \sqrt{n^2-1}/2$  of the physical mass $m $, the representation formula  allows  us also to derive a complete asymptotic expansion of the solution for the large time;
that is done in  Subsection~\ref{SS3.2cr}.

\section{The left knot point}
\label{S2}

The  equation  (\ref{25}) is strictly hyperbolic. That implies the well-posedness of the Cauchy problem for equation of (\ref{25})
in the various function spaces. The coefficient of the equation is an analytic function and, consequently, the
Holmgren's theorem  implies  local uniqueness in the space of distributions.
Moreover, the speed of propagation is finite, namely,
it is equal to $e^{-t} $ for every $ t \in {\mathbb R}$.
The second-order strictly hyperbolic  equation (\ref{25}) possesses two fundamental solutions
resolving the Cauchy problem. They can be written microlocally in terms of the Fourier integral operators \cite{Horm}, which
give a complete description of the wave front sets of the solutions.
The distance between two characteristic roots $\lambda_1 (t,\xi ) $ and $\lambda_2 (t,\xi ) $ of the equation (\ref{25}) is
$|\lambda_1 (t,\xi ) - \lambda_2 (t,\xi )| = e^{-t}|\xi |$, $ t \in {\mathbb R}$, $\xi \in {\mathbb R}^n$.
It tends to zero as $t $ approaches $\infty$. Thus, the operator is not uniformly  strictly hyperbolic.
Moreover, this equation possesses the   so-called {\it horizon}. More precisely,  any  signal emitted from the spatial point $x_0 \in {\mathbb R}^n$
at time $t_0 \in {\mathbb R} $ remains inside the ball $ |x -x_0 | < e^{-t_0} $
for all time $t \in ( t_0,\infty) $. 
\smallskip

By means  of the representation theorems for the solution of Cauchy's problem
we obtain in Section~\ref{SS3.1cr} a necessary and sufficient condition for the validity of
Huygens' principle  
for the field
equation  (\ref{25}). Huygens' principle plays an important role also
in quantum field theory in the curved spacetime. According to \cite{Lichnerowicz}
the support of the commutator-or the anticommutator-distribution, respectively,
lies on the null-cone if and only if Huygens' principle holds for the corresponding
wave equation.

Huygens' principle (or, more precisely, its ``minor premise'' due to
Hadamard) states that the support of the   
fundamental solution of a given hyperbolic   equation  
belongs to the surface of the characteristic conoid. In other words, the field equations (\ref{25})
satisfy Huygens' principle if and only if the solutions have no tail. Such domains in physical
spacetime wherein the fundamental distribution solution vanishes identically are referred to
as lacunas  of hyperbolic operators \cite{At'ja-Bott-Gording}. For the equation (\ref{25})
the complementary set of
the characteristic conoid consists of two open connected components.  The fact that
the exterior component is a lacuna proves the finiteness of the wave propagation
velocity. On the other hand, the existence of an inner lacuna, i.e. one that contains
time-like curves in the spacetime, is a very specific  property which
is intrinsic for a quite exceptional class of hyperbolic operators \cite{Friedlander,Gunther-book,Gunther}.

\smallskip

Consider now the knot points  for the physical mass,  $m^2= \frac{n^2}{4}-\left( \frac{1}{2} +k\right)^2$, $k=0,1,\ldots,\left[ \frac{n-1}{2}\right] $.
For $n=1$  only  the massless field $m=0$ has knot  point, 
while   for  $n=3$  there are two knot points. The knot points are linked to the Huygens' principle via intrinsic properties of the hypergeometric function.  
In fact, there  are  some  polynomials in the hierarchy of the  hypergeometric functions $ F \left(a, b;c; \zeta  \right)$. In particular,
 if $k \in {\mathbb N}$, then 
\[
  F \left(-k, -k;1; z  \right) =  \sum_{l=0}^{k}\left( \frac{k(k-1)\cdots (k+1-l)}{k!}\right)^2 z^{\,l}\,.
\]
For the corresponding $M$ we obtain $M=k+ \frac{1}{2}$, $k=0,1,\ldots,\left[ \frac{n-1}{2}\right] $.  If $ n $ is odd  and $k=\frac{n-1}{2}$, then we have  $M=\frac{n}{2} $.
Furthermore, for $M=\frac{3}{2} $  after simplifications, we obtain
\begin{eqnarray*} 
&  & 
E\left(x,t;x_0,t_0;\frac{3}{2}\right) \\ 
& = &
\frac{1}{8}  e^{ \frac{3}{2}(t_0+t) } \Big((e^{-t }+e^{-t_0})^2 - (x - x_0)^2\Big)  
F\Big(-1   ,-1  ;1;
\frac{ ( e^{-t_0}-e^{-t })^2 -(x- x_0 )^2 }{( e^{-t_0}+e^{-t })^2 -(x- x_0 )^2 } \Big) \\
& =  & 
\frac{1}{4}  e^{ \frac{3}{2}(t_0+t) }  \Big(   e^{-2t_0}  + e^{-2t }  - (x- x_0 )^2   \Big)
\end{eqnarray*}
while
\begin{eqnarray*} 
&  &
K_0\left(z,t;\frac{3}{2}\right)
  :=
- \left[  \frac{\partial }{\partial b}   E(z,t;0,b;M) \right]_{b=0} =
   \frac{1}{8}  e^{ \frac{3}{2}t } \left[ 3    ( z^2  - e^{-2t }    )
+     1        \right] 
\end{eqnarray*}
and
\begin{eqnarray*} 
&  &
K_1\left(z,t;\frac{3}{2}\right) = 
\frac{1}{4}e^{\frac{3}{2}t }  
 \left(   1+ e^{-2t }  -z^2  \right).   
 \end{eqnarray*}
For $M= \frac{3}{2}$, from (\ref{24}) we derive the following representation for the solution  
\begin{eqnarray*} 
\Phi  (x,t)
& = &
e^{-\frac{n-1}{2}t} v_{\varphi_0}  (x, \phi (t))\\
&  &
+ \,  \frac{1}{4}  e^{-\frac{n}{2}t+ \frac{3}{2}t}\int_{ 0}^{1} v_{\varphi_0}  (x, \phi (t)s) \\
&  &
\times \left(     3     \Big( (\phi (t)s)^2   - e^{-2t }  \Big)
+     1  + n   
 \left(   1+ e^{-2t }  -(\phi (t)s)^2  \right)\right)\phi (t)\,  ds  \nonumber \\
& &
+\,  \frac{1}{2} e^{-\frac{n}{2}t+ \frac{3}{2}t }\int_{0}^{\phi (t)}   v_{\varphi _1 } (x,  s)     
 \left(   1+ e^{-2t }  - s ^2  \right) \, ds, 
\quad x \in {\mathbb R}^n, \,\, t>0\,. \nonumber  
\end{eqnarray*} 
It can be rewritten as follows
\begin{eqnarray*} 
\Phi  (x,t)
& = &
e^{-\frac{n-1}{2}t} v_{\varphi_0}  (x, \phi (t))\\
&  &
+ \,  \frac{1}{4}  e^{-\frac{n}{2}t+ \frac{3}{2}t}\int_{ 0}^{\phi (t)} v_{\varphi_0}  (x,  s)\big(         (n -  3) e^{-2t } - (n-3)s ^2   
+     1  +   n     \big) \,  ds  \nonumber \\
& &
+\,  \frac{1}{2} e^{-\frac{n}{2}t+ \frac{3}{2}t }\int_{0}^{\phi (t)}   v_{\varphi _1 } (x,  s)     
 \left(   1+ e^{-2t }  - s ^2  \right) \, ds
, \quad x \in {\mathbb R}^n, \,\, t>0\,. \nonumber  
\end{eqnarray*}  
In particular, for $n=3$, consequently $m=0$, we obtain 
\begin{eqnarray} 
\Phi  (x,t)
& = &
e^{- t} v_{\varphi_0}  (x, \phi (t)) 
+ \,     \int_{ 0}^{\phi (t)} v_{\varphi_0}  (x,  s)  \,  ds  \nonumber \\
& &
+\,  \frac{1}{2}  \int_{0}^{\phi (t)}   v_{\varphi _1 } (x,  s)     
 \left(   1+ e^{-2t }  - s ^2  \right) \, ds
, \quad x \in {\mathbb R}^n, \,\, t>0\,. \nonumber  
\end{eqnarray}  
 Now, if we denote $V_{\varphi }$ the solution of
the problem 
\begin{eqnarray} 
\label{14}
V_{tt}-  \bigtriangleup V =0, \quad V(x,0)= 0, \quad V_t(x,0)=\varphi (x),
\end{eqnarray}
then
$
v_{\varphi }(x,t) =\frac{\partial }{\partial t}V_{\varphi }(x,t)\,,
$
and
\begin{eqnarray*} 
\Phi  (x,t)
& = &
e^{- t} v_{\varphi_0}  (x, \phi (t)) 
+ \,      V_{\varphi_0 }(x,\phi (t))  
+\,  \frac{1}{2} \left(   1+ e^{-2t }    \right)   V_{\varphi_1 }(x,\phi (t))   \\
&  & 
 -\,  \frac{1}{2}  \int_{0}^{\phi (t)}   v_{\varphi _1 } (x,  s)     
 s ^2  \, ds  \,. 
\end{eqnarray*} 
Hence, 
\begin{eqnarray*} 
\Phi  (x,t)
& = &
e^{- t} v_{\varphi_0}  (x, \phi (t)) 
+ \,      V_{\varphi_0 }(x,\phi (t))  
+\,  \frac{1}{2} \left(   1+ e^{-2t }    \right)   V_{\varphi_1 }(x,\phi (t))   \\
&  & 
 -\,  \frac{1}{2}V_{\varphi_1 }(x,\phi (t)) \phi^2 (t)  +\,  \int_{0}^{\phi (t)}   V_{\varphi _1 } (x,  s)     
 s    \, ds \,    
\end{eqnarray*} 
implies
\[ 
  \Phi  (x,t)
 =  
e^{- t} v_{\varphi_0}  (x, \phi (t)) 
+      V_{\varphi_0 }(x,\phi (t))  
+  e^{- t }   V_{\varphi_1 }(x,\phi (t))     + \int_{0}^{\phi (t)}   V_{\varphi _1 } (x,  s)     
 s   \,  ds .    
\]  
Thus, the sufficiency part of Theorem~\ref{TIHP} in the case of $n=3 $ is proven.
\medskip

Consider now the case of  $n=1$ and $M=\sqrt{\frac{1}{4}-m^2} $. There is only one knot point for such $n$ and $M$. Then we set $M= \frac{1}{2}$, consequently $m=0$, and obtain
\begin{eqnarray*} 
\Phi  (x,t)
& = &
 v_{\varphi_0}  (x, \phi (t))\\
&  &
+ \,  e^{-\frac{1}{2}t}\int_{ 0}^{1} v_{\varphi_0}  (x, \phi (t)s)\left(2  K_0 \left(\phi (t)s,t;\frac{1}{2}\right)+ K_1\left(\phi (t)s,t;\frac{1}{2}\right)\right)\phi (t)\,  ds  \nonumber \\
& &
+\, 2e^{-\frac{1}{2}t}\int_{0}^1   v_{\varphi _1 } (x, \phi (t) s)
  K_1\left(\phi (t)s,t;\frac{1}{2}\right) \phi (t)\, ds   \,. 
\end{eqnarray*} 
That is, the solution for the massless equation is given as follows
\begin{eqnarray*} 
\hspace{-0.4cm} \Phi  (x,t)
& = &
\frac{1}{2}\left( \varphi_0  (x-\phi (t))+  \varphi_0  (x+\phi (t))\right)  
+\frac{1}{2}\int_{0}^{\phi (t)}  \left( \varphi_1 (x-s)+  \varphi_1 (x+s)\right)\, ds  \,.
\end{eqnarray*} 
It also satisfies the  incomplete   Huygens' principle. The sufficiency part of Theorem~\ref{TIHP} is proven.

\section{Equation with the source term}
\label{Ssource}

Consider the linear part of the scalar equation
\begin{equation}
\label{K_G_Higgs}
u_{tt} - e^{-2t} \bigtriangleup u  - M^2   u=  - e^{\frac{n}{2}t}V'(e^{-\frac{n}{2}t}u ),
\end{equation}
with $M\geq 0 $. The  equation (\ref{K_G_Higgs}) includes two important cases. The first one is the Higgs boson equation, which has $V'(\phi )=\lambda \phi ^3 $
and $M^2=  \mu m^2+ n^2/4 $ with $\lambda >0 $ and $\mu >0 $, while $n=3$. The second case is for the small physical  mass, that is $0 \leq m  \le \frac{n }{2}$. For the last range of the mass we have
$ M^2= \frac{n^2}{4}-m^2$.
\smallskip

To prove the existence of the local and global solutions of the Cauchy problem for the equation (\ref{K_G_Higgs}) the useful tools are the representation 
formula for the solution of the linear equation with the source term and  some decay estimates for the norms of solution. We provide now the first one to complete the list of the representation formulas.   
The solution $u= u(x,t)$ to the Cauchy problem
\begin{eqnarray}
\label{eqim}
u_{tt} - e^{-2t}\Delta u -M^2 u= f ,\quad u(x,0)= 0  , \quad u_t(x,0)=0,
 \end{eqnarray}
with \, $ f \in C^\infty ({\mathbb R}^{n+1})$\, and with   vanishing
initial data is given in \cite{yagdjian_DCDS} by the next expression
\begin{eqnarray*}
u(x,t)
&  =  &
2   \int_{ 0}^{t} db
  \int_{ 0}^{ e^{-b}- e^{-t}} dr  \, \, v(x,r ;b) E(r,t; 0,b;M)  ,
\end{eqnarray*}
where the function
$v(x,t;b)$
is a solution to the Cauchy problem for the  wave equation:
\begin{equation}
\label{1.6} 
v_{tt} -   \bigtriangleup v  =  0 \,, \quad v(x,0;b)=f(x,b)\,, \quad v_t(x,0;b)= 0\,.
\end{equation}
The solution $u=u (x,t)$ to the Cauchy problem
\[
u_{tt}-  e^{-2t} \bigtriangleup u -M^2 u =0\,, \quad u(x,0)= \varphi_0 (x)\, , \quad u_t(x,0)=\varphi_1 (x)\,,
\]
with \, $ \varphi_0 $,  $ \varphi_1 \in C_0^\infty ({\mathbb R}^n) $, $n\geq 2$, can be represented (see \cite{yagdjian_DCDS}) as follows:
\begin{eqnarray*}
u(x,t)
& = &
 e ^{\frac{t}{2}} v_{\varphi_0}  (x, \phi (t))
+ \, 2\int_{ 0}^{1} v_{\varphi_0}  (x, \phi (t)s) K_0(\phi (t)s,t;M)\phi (t)\,  ds  \nonumber \\
& &
+\, 2\int_{0}^1   v_{\varphi _1 } (x, \phi (t) s)
  K_1(\phi (t)s,t;M) \phi (t)\, ds
, \quad x \in {\mathbb R}^n, \,\, t>0\,,
\end{eqnarray*}
where $\phi (t):=  1-e^{-t} $.
 Here, for $\varphi \in C_0^\infty ({\mathbb R}^n)$ and for $x \in {\mathbb R}^n$,
the function $v_\varphi  (x, \phi (t) s)$  coincides with the value $v(x, \phi (t) s) $
of the solution $v(x,t)$ of the Cauchy problem (\ref{1.11}).

Thus, for the solution $\Phi $ of the the Cauchy problem
\begin{eqnarray}
\label{17}
  \Phi _{tt} +   n   \Phi _t - e^{-2 t} \bigtriangleup \Phi  + m^2\Phi =  f ,\quad \Phi  (x,0)= 0  , \quad \Phi  _t(x,0)=0,
\end{eqnarray}
due to the relation $u = e^{\frac{n}{2}t}\Phi $, we obtain
with \, $ f \in C^\infty ({\mathbb R}^{n+1})$\, and with   vanishing
initial data   the next expression
\begin{eqnarray}
\label{Pfim}
\Phi  (x,t)
  =
2   e^{-\frac{n}{2}t}\int_{ 0}^{t} db
  \int_{ 0}^{ e^{-b}- e^{-t}} dr  \,  e^{\frac{n}{2}b} v(x,r ;b) E(r,t; 0,b;M)  ,
\end{eqnarray}
where the function
$v(x,t;b)$
is a solution to the Cauchy problem for the  wave equation (\ref{1.6}).

In fact, the representation formulas of this section have been used in \cite{Yag_CPDE_2012} to establish sign-changing 
properties of the global in time solutions of the Higgs boson equation.
\smallskip

\section{The right knot point. Proof of theorems}
\label{SS3.1cr}

Here we   
set $M=1/2$, that is, $m^2= (n^2-1)/4 $, which simplifies the hypergeometric functions, as well as, the kernels  $K_0 (z,t;M)$ and $K_1 (z,t;M)$. In that case we have

\begin{eqnarray*}
E\left(x,t;x_0,t_0;\frac{1}{2}\right) = 
 \frac{1}{2} e^{ \frac{1}{2}(t_0+t) } ,\quad E\left(z,t;0,b;\frac{1}{2}\right) = \frac{1}{2} e^{ \frac{1}{2}(b+t) }\,,
\end{eqnarray*}
while
\begin{eqnarray*}
K_0\left(z,t;\frac{1}{2} \right)
  =
- \frac{1}{4}  e^{ \frac{1}{2}t }  ,\qquad
K_1\left(z,t;\frac{1}{2} \right)
   =   \frac{1}{2}  e^{ \frac{1}{2}t }   \,.
 \end{eqnarray*}
For the solution (\ref{Pfim}) of the problem (\ref{17}) with the source term it follows
\begin{eqnarray*}
\Phi  (x,t)
&  =  & 
    e^{-\frac{n-1}{2}t}\int_{ 0}^{t}  e^{\frac{n+1}{2}b} db
  \int_{ 0}^{ e^{-b}- e^{-t}}v(x,r ;b)  \,   dr    \,,
\end{eqnarray*}
where the function $v(x,r ;b) $ is defined by (\ref{1.6}). In order to get rid of one integration in the last formula,  we denote $V_{f}(x,t;b)$ the solution of the problem
\begin{eqnarray*} 
V_{tt}-  \bigtriangleup V =0, \quad V(x,0)= 0, \quad V_t(x,0)=f (x,b)\,,
\end{eqnarray*}
then
\[
v(x,t;b) =\frac{\partial }{\partial t}V_{f}(x,t;b)\,.
\]
Hence,
\begin{eqnarray*}
\Phi  (x,t)
&  =  &
  e^{-\frac{n-1}{2}t}\int_{ 0}^{t}  e^{\frac{n+1}{2}b}
   V_{f}(x,e^{-b}- e^{-t};b)   \,   db .
\end{eqnarray*}
Further, due to  (\ref{24}) we have for the solution $\Phi $ of the  equation  without source term  the following representation 
\begin{eqnarray*}
\Phi  (x,t)
& = &
e^{-\frac{n-1}{2}t} v_{\varphi_0}  (x, 1-e^{-t})
+ \,   \frac{n-1}{2}e^{-\frac{n-1}{2}t}\int_{ 0}^{1-e^{-t}} v_{\varphi_0}  (x, s )        \,  ds  \nonumber \\
& &
+\,  e^{-\frac{n-1}{2}t}\int_{0}^{1-e^{-t}}   v_{\varphi _1 } (x, s )
     \, ds
, \quad x \in {\mathbb R}^n, \,\, t>0\,,
\end{eqnarray*}
where the functions $v_{\varphi_0 }    $ and $v_{\varphi _1 }  $ are defined by (\ref{1.11}). Now, if we denote $V_{\varphi }$ the solution of
the problem (\ref{14}), then
$
v_{\varphi }(x,t) =\frac{\partial }{\partial t}V_{\varphi }(x,t)\,,
$
and
\begin{eqnarray*}
\Phi  (x,t)
& = &
e^{-\frac{n-1}{2}t} v_{\varphi_0}  (x, 1-e^{-t})
+ \,   \frac{n-1}{2}e^{-\frac{n-1}{2}t}    V_{\varphi_0}  (x, 1-e^{-t} )    \nonumber \\
& &
+\,  e^{-\frac{n-1}{2}t}  V_{\varphi_1}  (x, 1-e^{-t} )
, \quad x \in {\mathbb R}^n, \,\, t>0\,    ,
\end{eqnarray*}
or, equivalently, 
\begin{eqnarray*}
\Phi  (x,t)
& = &
e^{-\frac{n-1}{2}t}  \left(  \frac{\partial V_{\varphi _0 }}{\partial t} \right) (x, 1-e^{-t})
+ \,   \frac{n-1}{2}e^{-\frac{n-1}{2}t}    V_{\varphi_0}  (x, 1-e^{-t} )    \nonumber \\
& &
+\,  e^{-\frac{n-1}{2}t}  V_{\varphi_1}  (x, 1-e^{-t} )
, \quad x \in {\mathbb R}^n, \,\, t>0\,.
\end{eqnarray*}
Thus,   we have proven the sufficiency  part of Theorem~\ref{T3}.  
\smallskip

Although the representation formulas make the proof of the necessity part  very clear and straightforward,   we provide  details of the proof in order 
to reveal the path that connects  the Huygens' principle with the values of mass $m$ and the dimension $n$. 
We   consider the case of small mass, $m \leq n/2$, since the relation between 
$E(x,t;x_0,t_0;M)$ and $E(x,t;x_0,t_0)$  (analytic continuation) shows the way how it can be proved  that, for the
large mass $ m >n/2$ the   Huygens' principle is not valid.

In order to prove the necessity of the conditions  $m=\sqrt{n^2-1}/2 $ and  $n$ is odd, we  set $M\not=\frac{1}{2}$, $\varphi _0=0$ and consider the solution (\ref{24})
of the Cauchy problem with the radial initial datum $\varphi _1  = \varphi _1  (r )$, supp$\,  \varphi _1 \subset \{ x\in {\mathbb R}^n\,;\,|x|\leq 1-\varepsilon \}$, $\varepsilon \in(0,1) $:
\begin{eqnarray*} 
\Phi  (x,t)
& = &
 2e^{-\frac{n}{2}t}\int_{0}^1   v_{\varphi _1 } (x, \phi (t) s)
  K_1(\phi (t)s,t;M) \phi (t)\, ds \\
& = &
 2e^{-\frac{n}{2}t}\int_{0}^ {\phi (t)}  \frac{\partial }{\partial s}V_{\varphi _1 }(x,s)  
  K_1( s,t;M)  \, ds\\
& = &
2e^{-\frac{n}{2}t}  V_{\varphi _1 }(x,\phi (t))  
  K_1( s,\phi (t);M) -
 2e^{-\frac{n}{2}t}\int_{0}^ {\phi (t)}  V_{\varphi _1 }(x,s)  
 \frac{\partial }{\partial s} K_1( s,t;M)  \, ds\,, 
\end{eqnarray*}
where   $V_{\varphi }$ is the solution of
the problem (\ref{14}) 
and $
v_{\varphi }(x,t) =\frac{\partial }{\partial t}V_{\varphi }(x,t)$.

 For odd $n$  and sufficiently large $t$ the point $(0,t)$  cannot be reached from the support of initial data by null geodesic. 
 The intersection of the support of  $\varphi _1   $ with the characteristic conoid $C_-(0,t)$ is empty, 
 and, consequently,   the contribution of the integral to the solution is crucial  for the validity of Huygens' principle. 
We consider the value of the solution at the spatial origin $x=0$:
\begin{eqnarray*} 
\Phi  (0,t)
& = & 
-  2e^{-\frac{n}{2}t}\int_{0}^ {\phi (t)}  V_{\varphi _1 }(0,s)  
 \frac{\partial }{\partial s} K_1( s,t;M)  \, ds\, \qquad \mbox{\rm for large}\quad t. 
\end{eqnarray*}
According to the well-known formula (see, e.g., \cite{Shatah}), we have 
\begin{eqnarray*} 
V_{\varphi_1 }(0, t) 
& =  &
\Big( \frac{1}{t} \frac{\partial }{\partial t}\Big)^{\frac{n-3}{2} }
\frac{t^{n-2}}{\omega_{n-1} c_0^{(n)} } \int_{S^{n-1}  }
\varphi_1 (ty)\, dS_y \\ 
& =  &
\left( \int_{S^{n-1}  }
\, dS_y \right) \frac{1}{\omega_{n-1} c_0^{(n)} } \Big( \frac{1}{t} \frac{\partial }{\partial t}\Big)^{\frac{n-3}{2} }
 t^{n-2}  \varphi_1 (t)\\ 
& =  &
\frac{1}{ c_0^{(n)} } \Big( \frac{1}{t} \frac{\partial }{\partial t}\Big)^{\frac{n-3}{2} }
 t^{n-2}  \varphi_1 (t),
\end{eqnarray*}
where $c_0^{(n)} =1\cdot 3\cdot \ldots \cdot (n-2)$ if $n \geq 3 $ is odd. Consequently, for large $t$ 
\begin{eqnarray*} 
\Phi  (0,t)
& = & 
 -  2e^{-\frac{n}{2}t}\int_{0}^ {\phi (t)}    \left[ \frac{1}{ c_0^{(n)} } \Big( \frac{1}{s} \frac{\partial }{\partial s}\Big)^{\frac{n-3}{2} }
 s^{n-2}  \varphi_1 (s) \right]
 \frac{\partial }{\partial s} K_1( s,t;M)  \, ds\\
 & = &
 -  2\frac{1}{ c_0^{(n)} }e^{-\frac{n}{2}t}\int_{0}^{1-\varepsilon }    \left[  \Big( \frac{1}{s} \frac{\partial }{\partial s}\Big)^{\frac{n-3}{2} }
 s^{n-2}  \varphi_1 (s) \right]
 \frac{\partial }{\partial s} K_1( s,t;M)  \, ds\,  .
\end{eqnarray*}

We evaluate the derivative 
$\frac{\partial }{\partial s} K_1( s,t;M)$:
\begin{eqnarray*} 
&  &
4^Me^{-Mt}\frac{\partial }{\partial s} K_1(s,t;M)  \\
& = &
2 \left(\frac{1}{2}-M\right) s \left(\left(1-e^{-t}\right)^2-s^2\right)^{-\frac{3}{2}+M} 
F\left(\frac{1}{2}-M,\frac{1}{2}-M;1;\frac{\left(1-e^{-t}\right)^2-s^2}{\left(1+e^{-t}\right)^2-s^2}\right)\\
&  &
-\left(\frac{1}{2}-M\right)^2 \left(\left(1-e^{-t}\right)^2-s^2\right)^{-\frac{1}{2}+M}
\frac{8 e^{3 t} s}{\left(1+ 2e^t+e^{2t} \left(1-s^2\right) \right)^2} \\
&  &
\times F\left(\frac{3}{2}-M,\frac{3}{2}-M;2;\frac{\left(1-e^{-t} \right)^2-s^2}{\left(1+e^{-t}\right)^2-s^2}\right)\\
& = &
 s \left(\left(1-e^{-t}\right)^2-s^2\right)^{-\frac{1}{2}+M}\left(\frac{1}{2}-M\right)\\
&  &
\times\Bigg\{  2 \left(\left(1-e^{-t}\right)^2-s^2\right)^{-1} 
F\left(\frac{1}{2}-M,\frac{1}{2}-M;1;\frac{\left(1-e^{-t}\right)^2-s^2}{\left(1+e^{-t}\right)^2-s^2}\right)\\
&  &
-\left(\frac{1}{2}-M\right) 
\frac{8 e^{3 t}}{\left(1+ 2e^t+e^{2t} \left(1-s^2\right) \right)^2}  F\left(\frac{3}{2}-M,\frac{3}{2}-M;2;\frac{\left(1-e^{-t} \right)^2-s^2}{\left(1+e^{-t}\right)^2-s^2}\right) \Bigg\}\,.
 \end{eqnarray*}
 Then, for the positive $M$  we have 
 \begin{eqnarray*} 
\lim_{t \to \infty }
e^{- t} F\left(\frac{3}{2}-M,\frac{3}{2}-M;2;\frac{\left(1-e^{-t} \right)^2-s^2}{\left(1+e^{-t}\right)^2-s^2}\right)=0\,,
 \end{eqnarray*}
while for $M=0$ we obtain
  \begin{eqnarray*} 
&  &
\lim_{z \to 1^- } (1-z) F\left(\frac{3}{2} ,\frac{3}{2}  ;2;z \right)= \frac{4}{\pi } \,, 
 \end{eqnarray*}   
 and, consequently,
\begin{eqnarray} 
\label{M12}
\lim_{t \to \infty }\frac{8 e^{3 t}}{\left(1+ 2e^t+e^{2t} \left(1-s^2\right) \right)^2}  F\left(\frac{3}{2},\frac{3}{2};2;\frac{\left(1-e^{-t} \right)^2-s^2}{\left(1+e^{-t}\right)^2-s^2}\right) 
& = &
\frac{8  }{\pi(1-s^2)}\,, 
 \end{eqnarray}
uniformly with respect to  $s\in [0,1-\varepsilon ]$.  
 
According to Subsection 2.1.3~\cite{B-E} if $\Re (c-a-b)>0$, then $F  ( a,b;c;1  ) = \frac{ \Gamma (c)\Gamma (c-a-b) }{\Gamma (c-a)\Gamma (c-b)}$, where $\Gamma  $ is the gamma-function.  For the positive $M$ such that $ M \not=\frac{1}{2} $, that implies
 \begin{eqnarray*} 
&  &
\lim_{t \to +\infty } F\left(\frac{1}{2}-M,\frac{1}{2}-M;1;\frac{\left(1-e^{-t}\right)^2-s^2}{\left(1+e^{-t}\right)^2-s^2}\right)= \frac{ \Gamma (2M) }{(\Gamma (\frac{1}{2} +M))^2} \,.
 \end{eqnarray*} 
Hence, for   $M>0$ it follows (See 15.3.6 of Ch.15\cite{A-S}  and \cite{B-E}.) 
\begin{eqnarray*} 
&  &
\lim_{t \to + \infty } 4^Me^{-Mt}\frac{\partial }{\partial s} K_1(s,t;M) \\
& = &
\lim_{t \to + \infty }  s \left(\left(1-e^{-t}\right)^2-s^2\right)^{-\frac{1}{2}+M}\left(\frac{1}{2}-M\right)\\
&  &
\times\Bigg\{  2 \left(\left(1-e^{-t}\right)^2-s^2\right)^{-1} 
F\left(\frac{1}{2}-M,\frac{1}{2}-M;1;\frac{\left(1-e^{-t}\right)^2-s^2}{\left(1+e^{-t}\right)^2-s^2}\right)\\
&  &
-\left(\frac{1}{2}-M\right) 
\frac{8 e^{3 t}}{\left(1+ 2e^t+e^{2t} \left(1-s^2\right) \right)^2}  F\left(\frac{3}{2}-M,\frac{3}{2}-M;2;\frac{\left(1-e^{-t} \right)^2-s^2}{\left(1+e^{-t}\right)^2-s^2}\right) \Bigg\}\\
& = &
\left(\frac{1}{2}-M\right) 2  
\frac{ \Gamma (2M) }{(\Gamma (\frac{1}{2} +M))^2} s \left(1-s^2\right)^{-\frac{3}{2}+M}\,,
 \end{eqnarray*}
uniformly with respect to  $s \in [0,1-\varepsilon ]$.  Hence, for the positive $M$ one can write  
\begin{eqnarray*} 
\lim_{t \to + \infty } 4^Me^{-Mt}\frac{\partial }{\partial s} K_1(s,t;M)
& = &
-2 \left(M-\frac{1}{2} \right)
\frac{\Gamma (2M)}{(\Gamma (\frac{1}{2} +M))^2}  s (1 - s^2)^{M-\frac{3}{2}  } \,.
 \end{eqnarray*}
 The last equation implies  that derivative is a sign preserving function in $(0,1)$. For $M>0$, $M\not= \frac{1}{2}$, the derivative vanishes for all $s \in (0,1)$ if and only if $M=\frac{1}{2}$, that is, 
 \begin{eqnarray*} 
\lim_{t \to + \infty } 4^Me^{-Mt}\frac{\partial }{\partial s} K_1(s,t;M)
& \not= &
0 \quad \mbox{for all}  \quad s\in (0,1-\varepsilon )\,.
 \end{eqnarray*} 
 In particular,
 \begin{eqnarray*} 
&  &
\lim_{t \to + \infty }\int_{0}^{1-\varepsilon }    \left[  \Big( \frac{1}{s} \frac{\partial }{\partial s}\Big)^{\frac{n-3}{2} }
 s^{n-2}  \varphi_1 (s) \right]
 4^Me^{-Mt}\frac{\partial }{\partial s} K_1( s,t;M)  \, ds  \\
 & = &
 \int_{0}^{1-\varepsilon }    \left[  \Big( \frac{1}{s} \frac{\partial }{\partial s}\Big)^{\frac{n-3}{2} }
 s^{n-2}  \varphi_1 (s) \right]
 \lim_{t \to + \infty }4^Me^{-Mt}\frac{\partial }{\partial s} K_1( s,t;M)  \, ds \\
 & = &
-2 \left(M-\frac{1}{2} \right) 
\frac{\Gamma (2M)}{(\Gamma (\frac{1}{2} +M))^2}\int_{0}^{1-\varepsilon }    \left[  \Big( \frac{1}{s} \frac{\partial }{\partial s}\Big)^{\frac{n-3}{2} }
 s^{n-2}  \varphi_1 (s) \right]
 s (1 - s^2)^{M-\frac{3}{2}  } \, ds \,.
\end{eqnarray*}
 Consequently,
\begin{eqnarray*} 
&  &
\lim_{t \to + \infty }\int_{0}^ {\phi (t)}  V_{\varphi _1 }(0,s)  
4^Me^{-Mt} \frac{\partial }{\partial s} K_1( s,t;M)  \, ds   \\
& = &
- 2\left(M-\frac{1}{2} \right)\frac{\Gamma (2M)}{(\Gamma (\frac{1}{2} +M))^2c_0^{(n)} } \int_{0}^ {1} s (1 - s^2)^{M-\frac{3}{2}  }   \Big( \frac{1}{s} \frac{\partial }{\partial s}\Big)^{\frac{n-3}{2} }
s^{n-2}  \varphi_1 (s)  
 \, ds \,.
 \end{eqnarray*}
 Hence,
\begin{eqnarray*} 
&   & 
4^{-M}e^{Mt} e^{-\frac{n}{2}t}\int_{0}^ {\phi (t)}  V_{\varphi _1 }(0,s)  
4^Me^{-Mt} \frac{\partial }{\partial s} K_1( s,t;M)  \, ds\\
&= & 
4^{-M}e^{Mt} e^{-\frac{n}{2}t}\left\{ \int_{0}^ {\phi (t)}  V_{\varphi _1 }(0,s)  
4^Me^{-Mt} \frac{\partial }{\partial s} K_1( s,t;M)  \, ds \right.\\
&  &
\left.- \lim_{t \to + \infty }\int_{0}^ {\phi (t)}  V_{\varphi _1 }(0,s)  
4^Me^{-Mt} \frac{\partial }{\partial s} K_1( s,t;M)  \, ds\right\}\\
& & 
+ 4^{-M}e^{Mt} e^{-\frac{n}{2}t}\left\{ \lim_{t \to + \infty }\int_{0}^ {\phi (t)}  V_{\varphi _1 }(0,s)  
4^Me^{-Mt} \frac{\partial }{\partial s} K_1( s,t;M)  \, ds\right\}\\
&= & 
4^{-M}e^{Mt} e^{-\frac{n}{2}t}\left\{o(1) 
+  \lim_{t \to + \infty }\int_{0}^ {\phi (t)}  V_{\varphi _1 }(0,s)  
4^Me^{-Mt} \frac{\partial }{\partial s} K_1( s,t;M)  \, ds\right\}\,, 
\end{eqnarray*}
where $ o(1) \to 0$ as $t \to \infty$. Finally,
\begin{eqnarray*} 
&   & 
 e^{-\frac{n}{2}t}\int_{0}^ {\phi (t)}  V_{\varphi _1 }(0,s)  
 \frac{\partial }{\partial s} K_1( s,t;M)  \, ds = 
4^{-M}e^{Mt}e^{-\frac{n}{2}t}\Bigg\{o(1) \\
  &  &
 - 2\left(M-\frac{1}{2} \right)\frac{\Gamma (2M)}{(\Gamma (\frac{1}{2} +M))^2c_0^{(n)} } \int_{0}^ {1} s (1 - s^2)^{M-\frac{3}{2}  }   \Big( \frac{1}{s} \frac{\partial }{\partial s}\Big)^{\frac{n-3}{2} }
s^{n-2}  \varphi_1 (s)  
 \, ds \Bigg\}\,. 
\end{eqnarray*}
 The last equation shows that for positive $M$, $M\not=1/2$, the value $ \Phi (0,t)$  of the solution   $\Phi =\Phi (x,t)$ for large $t$  depends on the values of the initial data inside of the characteristic conoid.

The case of $M=0$ can be discussed in similar way if we take into account (\ref{M12}), the support of the function $ V_{\varphi _1 }(0,s) $,  and 
\begin{eqnarray*} 
&  &
\lim_{t \to \infty} \frac{1}{\ln \left(1-\frac{\left(1-e^{-t}\right)^2-s^2}{\left(1+e^{-t}\right)^2-s^2}\right)}F\left(\frac{1}{2},\frac{1}{2};1;\frac{\left(1-e^{-t}\right)^2-s^2}{\left(1+e^{-t}\right)^2-s^2}\right)= -\frac{1}{\pi}\,.
 \end{eqnarray*}
 If $n$ is even, then the violation of the Huygens' principle is inherited from the Minkowski spacetime through  the representation formula. We skip the details of the proof of that case.  Theorem~\ref{T3} is proven. 
\hfill $\square$

\bigskip

\noindent
{\bf Proof of Theorem~\ref{TIHP}}. The arguments have been used in the proof of Theorem~\ref{T3} help us to prove also Theorem~\ref{TIHP}.
In order to exclude the equations that obey the Huygens' principle, we set $M \not= \frac{1}{2} $. Then we consider odd $n$, set    $ \varphi _1=0$ and   
choose the radial function $\varphi _0  = \varphi _0  (r )$, supp$\,  \varphi _0 \subset \{ x\in {\mathbb R}^n\,;\,|x|<1-\varepsilon \}$, $\varepsilon \in (0,1) $. 
The solution (\ref{24}) of the Cauchy problem is the following function
\begin{eqnarray*} 
\Phi  (x,t)
& = &
e^{-\frac{n-1}{2}t} v_{\varphi_0}  (x, \phi (t))\\
&  &
+ \,  e^{-\frac{n}{2}t}\int_{ 0}^{\phi (t)} v_{\varphi_0}  (x, s)\big(2  K_0(s,t;M)+ nK_1(s,t;M)\big)\,  ds 
, \quad x \in {\mathbb R}^n, \,\, t>0\,. \nonumber 
\end{eqnarray*}
To complete the proof of theorem it remains to find the principal term of the asymptotic of   the derivative
\begin{eqnarray*} 
  \frac{\partial }{\partial s} \big(2  K_0(s,t;M)+ nK_1(s,t;M)\big) =  2  \frac{\partial }{\partial s}K_0(s,t;M)+ n\frac{\partial }{\partial s}K_1(s,t;M) 
\end{eqnarray*}
for large $t$ and for $s \in [0,1-\varepsilon]$ on the support of $v_{\varphi_0}  (0, \cdot ) $. The second term of the right-hand side  of the derivative is already discussed above. We evaluate the first term:
\begin{eqnarray*}
&  &
 \frac{\partial }{\partial s}K_0(s,t;M)
  \\
&  = &
4 ^ {-M}  e^{ t M} \frac{\partial }{\partial s} \Bigg\{ \big((1+e^{-t })^2 - s^2\big)^{  M    } \frac{1}{ [(1-e^{ -t} )^2 -  s^2]\sqrt{(1+e^{-t } )^2 - s^2} }
\nonumber \\
&   &
\times  \Bigg[  \big(  e^{-t} -1 +M(e^{ -2t} -      1 -  s^2) \big)
F \Big(\frac{1}{2}-M   ,\frac{1}{2}-M  ;1; \frac{ ( 1-e^{-t })^2 -s^2 }{( 1+e^{-t })^2 -s^2 }\Big)   \\
&  &
\hspace{0.3cm}  +   \big( 1-e^{-2 t}+  z^2 \big)\Big( \frac{1}{2}+M\Big)
F \Big(-\frac{1}{2}-M   ,\frac{1}{2}-M  ;1; \frac{ ( 1-e^{-t })^2 -s^2 }{( 1+e^{-t })^2 -s^2 }\Big) \Bigg] \Bigg\}.
\end{eqnarray*}
On the other hand, for the positive $M$ the equation
\begin{eqnarray*}
&  &
F\left(-\frac{1}{2}-M,\frac{1}{2}-M;1;1\right)= \frac{4 M}{1+2 M}{ F\left(\frac{1}{2}-M,\frac{1}{2}-M;1;1\right)}
\end{eqnarray*}
implies
\begin{eqnarray*}
&  &
\lim_{t \to + \infty } 2^{1+2 M}e^{-Mt}\frac{\partial }{\partial s}K_0(s,t;M)
  \\
&  = &
\lim_{t \to + \infty }
\frac{1}{\left(1-s^2\right)^2} e^{-4 t}  s \left(1-s^2\right)^{-\frac{3}{2}+M}\\
&  &
\times  \Bigg\{ -(1+2 M) e^{4 t} \left(-7+6 s^2+s^4\right)  F\left(-\frac{1}{2}-M,\frac{1}{2}-M;1;1\right) \\
&  &
+2 \Big(M+2 M^2+e^{3 t} \left(4 M^2-1\right) \left(1+s^2\right)\\
&  &
+e^{4 t} (1+2 M) \left(-1+s^2\right) \left(3+M+M s^2\right)-e^{2 t} (1+2 M) \left(-3+2 M s^2\right)\Big) \\
&  &
\times F\left(\frac{1}{2}-M,\frac{1}{2}-M;1;1\right)\Bigg\} \\
&  = &
(1+2 M)\lim_{t \to + \infty }
\frac{1}{\left(1-s^2\right)^2} e^{-4 t}  s \left(1-s^2\right)^{-\frac{3}{2}+M}\\
&  &
\times  \Bigg\{ - e^{4 t} \left(-7+6 s^2+s^4\right)  F\left(-\frac{1}{2}-M,\frac{1}{2}-M;1;1\right) \\
&  &
+2 e^{4 t} \left(-1+s^2\right) \left(3+M+M s^2\right) F\left(\frac{1}{2}-M,\frac{1}{2}-M;1;1\right)\Bigg\}.
\end{eqnarray*}
It follows
\begin{eqnarray*}
&  &
\lim_{t \to + \infty } 2^{1+2 M}e^{-Mt}\frac{\partial }{\partial s}K_0(s,t;M)
  \\
&  = &
   s \left(1-s^2\right)^{-\frac{7}{2}+M}\left\{ \left(7-6 s^2- s^4\right)4 M 
+2 (1+2 M)(s^2-1) \left(3+M+M s^2\right)\right\}\\
&  &
\times   F\left(\frac{1}{2}-M,\frac{1}{2}-M,1,1\right).
\end{eqnarray*}
Consequently,
\begin{eqnarray*} 
&  &
 \lim_{t \to + \infty } 2^{1+2 M}e^{-Mt} \frac{\partial }{\partial s} \big(2  K_0(s,t;M)+ nK_1(s,t;M)\big) \\
& = &
 2   s \left(1-s^2\right)^{-\frac{7}{2}+M}\Big\{ (7-6 s^2- s^4)4 M 
+2 (1+2 M)(s^2-1) \left(3+M+M s^2\right)\Big\} \\
&  &
\times  F\left(\frac{1}{2}-M,\frac{1}{2}-M,1,1\right)\\
&  &
- 4n \left(M-\frac{1}{2} \right) s (1 - s^2)^{M-\frac{3}{2}  }
F\left(\frac{1}{2} - M, \frac{1}{2} - M; 1; 1\right)\\
& = &
\left(1-s^2\right)^{-\frac{5}{2}+M} s \Big\{  (7+ s^2 )8 M 
-4 (1+2 M) \left(3+M+M s^2\right)  
- 4n \left(M-\frac{1}{2} \right) (1 - s^2)\Big\}\\
&  &
\times F\left(\frac{1}{2} - M, \frac{1}{2} - M; 1; 1\right)\\
& = &
 -8 \left(1-s^2\right)^{-\frac{5}{2}+M} s  \left(M-\frac{1}{2}\right) \left( s^2\left(M-\frac{n}{2}\right)+ M+\frac{n}{2}  - 3  \right) F\left(\frac{1}{2} - M, \frac{1}{2} - M; 1; 1\right).
\end{eqnarray*}
The factor  
\[
 \left(M-\frac{1}{2}\right) \left( s^2\left(M-\frac{n}{2}\right)+ M+\frac{n}{2}  - 3  \right)
\]
with $M\not= \frac{1}{2} $   identically vanishes only if  $M=n/2$ and $n=3$. The rest of the proof is a repetition  of the one has been done above.
  The case of large mass, $m\geq  n/2$ can be checked similarly. 
 Theorem~\ref{TIHP} is proven. \hfill $\square$

\subsection{Asymptotic expansions of solutions   at infinite time}
\label{SS3.2cr}

In this subsection we present  the large time  asymptotic analysis of the solution of the equation, which obeys the Huygens' principle. 
More precisely, we derive the complete asymptotic expansion.  In fact, this analysis was started in the previous subsection.
\smallskip

Concerning asymptotic expansion of the solution for the large time, we mention here two recent 
articles on linear equations on the asymptotically de~Sitter spacetimes. Vasy~\cite{Vasy_2010}
exhibited the well-posedness of the Cauchy problem and showed that on such spaces, the solution
of the Klein-Gordon equation without source term  and with smooth Cauchy data has an asymptotic expansion at infinity.  It is  also shown in \cite{Vasy_2010}  
that the solutions of the wave equation exhibit scattering. Baskin~\cite{Baskin} constructed parametrix for the forward fundamental 
solution of the wave and Klein-Gordon equations on asymptotically de Sitter spaces without caustics and used this parametrix to obtain
asymptotic expansions (principal term)  for the solutions of the equation with some class of source terms. 
(For more references on the asymptotically de Sitter spaces, see  the bibliography in \cite{Baskin},
 \cite{Vasy_2010}.) 
\smallskip

For $\varphi_1 \in C_0^\infty ({\mathbb R}^n)$ 
the formula  for the solution $V (x, t)$   of the Cauchy problem (\ref{14})
is well-known. (See, e.g.,\cite{Shatah}.)  It can be written for odd and even $ n$ separately as follows. We have 
\begin{eqnarray*} 
V_{\varphi }(x, t) :=
 \Big( \frac{1}{t} \frac{\partial }{\partial t}\Big)^{\frac{n-3}{2} }
\frac{t^{n-2}}{\omega_{n-1} c_0^{(n)} } \int_{S^{n-1}  }
\varphi (x+ty)\, dS_y ,
\end{eqnarray*}
where $c_0^{(n)} =1\cdot 3\cdot \ldots \cdot (n-2)$ if $n \geq 3 $ is odd. 
For $x \in {\mathbb R}^n$, and even $n$,    we have
\begin{eqnarray*} 
V_{\varphi } (x, t) :=
\Big( \frac{1}{t} \frac{\partial }{\partial t}\Big)^{\frac{n-2}{2} }
\frac{2t^{n-1}}{\omega_{n-1} c_0^{(n)}} \int_{B_1^{n}(0)}  \frac{1}{\sqrt{1-|y|^2}}\varphi  (x+ty)\, dV_y \,,
\end{eqnarray*}
where $c_0^{(n)} =1\cdot 3\cdot \ldots \cdot (n-1)$.
Similarly, for $\varphi_0 \in C_0^\infty ({\mathbb R}^n)$ and for $x \in {\mathbb R}^n$, if $n $ is odd,  
the formula for the solution $u  (x, t)$   of the Cauchy problem
\[
u_{tt}-  \Delta  u =0, \quad u(x,0)= \varphi_0 (x), \quad u_t(x,0)=0\,,
\]
implies 
\begin{eqnarray*} 
v_{\varphi }   (x, t) :=
  \frac{\partial}{\partial t} \Big( \frac{1}{t} \frac{\partial }{\partial t}\Big)^{\frac{n-3}{2} }
\frac{t^{n-2}}{\omega_{n-1} c_0^{(n)} } \int_{S^{n-1}  }
\varphi (x+ty)\, dS_y \,.
\end{eqnarray*}
In the case of 
  $x \in {\mathbb R}^n$ and even $n $,   we have
\begin{eqnarray*} 
v_{\varphi }  (x, t) :=    \frac{\partial }{\partial t}
\Big( \frac{1}{t} \frac{\partial }{\partial t}\Big)^{\frac{n-2}{2} }
\frac{2t^{n-1}}{\omega_{n-1} c_0^{(n)}} \int_{B_1^{n}(0)}  
\frac{1}{\sqrt{1-|y|^2}}\varphi (x+ty)\, dV_y   \,.
\end{eqnarray*} 
The constant $\omega_{n-1} $ is the area of the unit sphere $S^{n-1} \subset {\mathbb R}^n$. 
In particular,
\begin{eqnarray*}
v_{\varphi }   (x, 1) 
& = &
\cases{   \left[ \frac{\partial}{\partial t} \Big( \frac{1}{t} \frac{\partial }{\partial t}\Big)^{\frac{n-3}{2} }
\frac{t^{n-2}}{\omega_{n-1} c_0^{(n)} } \int_{S^{n-1}  }
\varphi (x+ty)\, dS_y \right]_{t=1}, \quad   \mbox{\rm if}\,\,  n \,\, \mbox{\rm is odd,}\cr 
\left[   \frac{\partial }{\partial t}
\Big( \frac{1}{t} \frac{\partial }{\partial t}\Big)^{\frac{n-2}{2} }
\frac{2t^{n-1}}{\omega_{n-1} c_0^{(n)}} \int_{B_1^{n}(0)}  
\frac{1}{\sqrt{1-|y|^2}}\varphi (x+ty)\, dV_y  \right]_{t=1}, \quad  \mbox{\rm if}\,\,  n \,\, \mbox{\rm is even,}
}
\end{eqnarray*}
and 
\begin{eqnarray*}
V_{\varphi }(x, 1) 
& = &
\cases{   
 \left[ \Big( \frac{1}{t} \frac{\partial }{\partial t}\Big)^{\frac{n-3}{2} }
\frac{t^{n-2}}{\omega_{n-1} c_0^{(n)} } \int_{S^{n-1}  }
\varphi (x+ty)\, dS_y ,\right]_{t=1}, \quad  \mbox{\rm if}\,\,  n \,\, \mbox{\rm is odd,}\cr 
\left[ \Big( \frac{1}{t} \frac{\partial }{\partial t}\Big)^{\frac{n-2}{2} }
\frac{2t^{n-1}}{\omega_{n-1} c_0^{(n)}} \int_{B_1^{n}(0)}  \frac{1}{\sqrt{1-|y|^2}}\varphi  (x+ty)\, dV_y \right]_{t=1}, \quad  \mbox{\rm if}\,\, n \,\, \mbox{\rm is even.}
 }
\end{eqnarray*}
Denote
\begin{eqnarray*}
v_{\varphi }   (x):= v_{\varphi }   (x, 1)\,, \qquad V_{\varphi }(x):=V_{\varphi }(x, 1)\,.
\end{eqnarray*}
In order to write complete asymptotic expansion of the solutions,   we define the functions
\begin{eqnarray*}
V_{\varphi }^{(k)}(x) 
& = &
\frac{(-1)^k}{k!}\left[\left( \frac{\partial }{\partial t} \right)^k V_{\varphi }(x, t) \right]_{t=1}
\in C_0^\infty ({\mathbb R}^n)\,,\quad k=1,2,\ldots\,.
\end{eqnarray*}
Then, for every integer  $N \geq 1$ we have
\begin{eqnarray*}
V_{\varphi }(x,1-e^{-t}) = \sum_{k=0}^{N-1} V_{\varphi }^{(k)}(x)e^{-kt} + R_{V_\varphi ,N}(x,t),  \quad   
R_{V_\varphi ,N} \in   C ^\infty \,, 
\end{eqnarray*}
where  the remainder  $ R_{V_\varphi ,N} $ satisfies the inequality 
\begin{eqnarray*}
| R_{V_\varphi ,N}(x,t) | \leq C(\varphi ) e^{-Nt}\quad  \,\, \mbox{\rm for all}\quad  x \in {\mathbb R}^n \quad  \mbox{\rm and all}\quad  t \in [0,\infty)  \,, 
\end{eqnarray*}
with some constant $ C(\varphi ) $. Moreover, the support of the remainder $ R_{V_\varphi ,N} $ is in the cylinder 
\begin{eqnarray*}
 \mbox{\rm supp}\,   R_{V_\varphi ,N}   \subseteq 
   \{ x \in {\mathbb R}^n \,;\, \mbox{\rm dist} (x,   \mbox{\rm supp}\, \varphi ) \leq 1\, \}\times [0,\infty)  \,. 
\end{eqnarray*}
Analogously, we define
\begin{eqnarray*}
v_{\varphi }^{(k)}(x) 
& = &
\frac{(-1)^k}{k!}\left[\left( \frac{\partial }{\partial t} \right)^k v_{\varphi }(x, t) \right]_{t=1}
\in C_0^\infty ({\mathbb R}^n)\,,\quad k=1,2,\ldots\,,
\end{eqnarray*}
and the remainder $ R_{v_\varphi ,N} $,  
\begin{eqnarray*}
v_{\varphi }(x,1-e^{-t}) = \sum_{k=0}^{N-1} v_{\varphi }^{(k)}(x)e^{-kt} + R_{v_\varphi ,N}(x,t),  \qquad   
R_{v_\varphi ,N} \in   C ^\infty \,, 
\end{eqnarray*}
such that
\begin{eqnarray*}
| R_{v_\varphi ,N}(x,t) | \leq C(\varphi ) e^{-Nt}\quad  \,\, \mbox{\rm for all}\quad  x \in {\mathbb R}^n \quad  \mbox{\rm and all}\quad  t \in [0,\infty)  \,. 
\end{eqnarray*}
Further,  we introduce  a polynomial in $z  \in {\mathbb C} $ with the smooth in $x \in {\mathbb R}^n$ coefficients  as follows:
\[ 
\Phi_{asypt}^{(N)}  (x,z) 
 = 
 z^{\frac{n-1}{2}}  \left(    \sum_{k=0}^{N-1} v_{\varphi_0 }^{(k)}(x)z^k   
+     \frac{n-1}{2}     \sum_{k=0}^{N-1} V_{\varphi_0 }^{(k)}(x)z^k  \right) 
+     z^{\frac{n-1}{2}} \sum_{k=0}^{N-1} V_{\varphi_1 }^{(k)}(x)z^k    \,.
\]
 Then we    write the next asymptotic expansion 
\begin{eqnarray*}
 \Phi  (x,t) = \Phi_{asypt}^{(N)}  (x,e^{-t})  + O(e^{-Nt-\frac{n-1}{2}t}) 
\end{eqnarray*}
for large $t$ uniformly for all $x\in {\mathbb R}^n$. Thus, we have proven the next theorem.

\begin{theorem}
\label{Tasymp}
Suppose that  $m= \sqrt{ n^2-1 }/2 $. Then, for every positive integer $N$   the solution of the Cauchy problem for the equation (\ref{25}) with the initial values $ \varphi_0, \varphi _1  \in C_0^\infty ({\mathbb R}^n) $   has the following asymptotic expansion at infinity:
\begin{eqnarray*}
\Phi  (x,t) \sim  \Phi_{asypt}^{(N)}  (x,e^{-t})\,,
\end{eqnarray*}
in the sense that  for every positive integer $N$  the following    estimate is valid,
\begin{eqnarray*}
\| \Phi  (x,t) - \Phi_{asypt}^{(N)}  (x,e^{-t}) \|_{L^\infty  ({\mathbb R}^n)}
& \leq  & 
C(\varphi_0, \varphi _1) e^{-Nt-\frac{n-1}{2}t } \quad \,\, \mbox{  for large }\,\, t\,.
\end{eqnarray*}
\end{theorem}
\smallskip

\begin{remark}
If we take into account the relation 
$v_{\varphi }   (x, t) = \frac{\partial }{\partial t} V_{\varphi } (x, t) $, then 
\begin{eqnarray*}
&  &
v_{\varphi }^{(k)}   (x) = - (k+1)  V_{\varphi }^{(k+1)}   (x)\,,
\end{eqnarray*}
and, consequently, the function $\Phi_{asypt}^{(N)}  (x,z) $ can be rewritten as follows:
\begin{eqnarray*}
&  &
\Phi_{asypt}^{(N)}  (x,z) \\
& = & 
 z^{\frac{n-1}{2}}  \left(    \sum_{k=0}^{N-1} (-1)(k+1) V_{\varphi_0 }^{(k+1)}(x)z^k   
+     \frac{n-1}{2}     \sum_{k=0}^{N-1} V_{\varphi_0 }^{(k)}(x)z^k  \right) 
+     z^{\frac{n-1}{2}} \sum_{k=0}^{N-1} V_{\varphi_1 }^{(k)}(x)z^k   \\ 
& = & 
 z^{\frac{n-1}{2}}    \sum_{k=0}^{N-1}\left( \frac{n-1}{2}      V_{\varphi_0 }^{(k)}(x)  -(k+1) V_{\varphi_0 }^{(k+1)}(x)   
+   V_{\varphi_1 }^{(k)}(x)\right) z^k .
\end{eqnarray*}
\end{remark}
In the forthcoming paper we will derive a similar result for the remaining  values of the mass $ m \in [0,\infty)$, that is, for the equation, which does not obey the Huygens' principle.

\end{document}